\documentclass[twocolumn,secnumarabic,amssymb, nobibnotes, aps, pra,showpacs,preprintnumbers]{revtex4-1}
\usepackage{amssymb}
\usepackage{array}
\usepackage{graphicx}

\def\beq{\begin{equation}}
\def\eeq{\end{equation}}
\def\bea{\begin{eqnarray}}
\def\eea{\end{eqnarray}}
\def\no{\nonumber}

\begin{document}

\title{Correspondence of phase transition points and singularities of thermodynamic geometry of black holes}%

\author{Seyed Ali Hosseini Mansoori}
\email{sa.hosseinimansoori@ph.iut.ac.ir}
\author{Behrouz Mirza}
\email{b.mirza@cc.iut.ac.ir}
\affiliation{Department of Physics,
Isfahan University of Technology, Isfahan 84156-83111, Iran}

\date{\today}%

\begin{abstract}
We explore a formulation of  thermodynamic geometry of black holes and prove that the divergent points of the specific heat correspond exactly to the singularities of the thermodynamic curvature. We investigate  this correspondence for different types of black holes. This formulation can also be applied to an arbitrary thermodynamic system.
\end{abstract}


\maketitle

\section{Introduction}\label{a}
 In 1971, Stephen Hawking \cite{L1} stated that the area, $A$, of the event horizon
of a black hole can never decrease in physical processes. It was later noted by Bekenstein \cite{L2} that this result was analogous to the statement of the ordinary second law of thermodynamics, namely that the total entropy, $S$, of a closed system never decreases. Bekenstein proposed that the entropy of a black hole is proportional to its area. Correspondence between thermodynamics of black holes and the known first and second laws of thermodynamics was furthur studied in \cite{aman4}. However, it is still a challenging problem to find the statistical origin of black hole thermodynamics.

The geometric formulation of thermodynamics is a useful tool to study some aspects of physical systems. For instance, Weinhold \cite{tun1} introduced in the equilibrium space a Riemannian metric defined in terms of the second derivatives of the internal energy with respect to entropy and other extensive variables of a thermodynamic system. Moreover, in 1979, Ruppeiner \cite{tun2} introduced a Riemannian metric structure in thermodynamic fluctuation theory, and related it to the second derivatives of the entropy. The Ruppeiner metric is based on the thermodynamic state space geometry. For the second order phase transitions, the Ruppeiner scalar curvature  (R) is expected to diverge at the critical point \cite{rup,brody,fff,mirza1,Lala1,Rabin1,Az1}. Over the last decade, thermodynamic geometry and some of its new formulations have also been applied to black holes
 \cite{Ferara,Aman,Shen,Mirza,Medved,Rup3,Man1}.
Another geometric formulation of thermodynamics was proposed by Quvedo \cite{quved}.  Quvedo's method incorporates Legendre invariance in a natural way, and allows us to derive Legendre invariant metrics in the space of equilibrium states. However,  this method not only contains certain ambiguities but also
 fails to explain the correspondence between phase transitions and singularities of the scalar curvature for Phantom Reissner-Nordstrom-AdS black hole \cite{cardy,R1,nettele1}.

In this paper, we  explore a formulation for the thermodynamic geometry of black holes, and prove that the divergent points of the specific heat correspond exactly to the singularities of the thermodynamic geometry.
The outline of this paper is as follows. In Section 2,  certain analytical techniques are used to  prove that  singularities of specific heat and scalar curvature occur on identical points. In Section 3, we study the thermodynamic geometry of the phantom Reissner- Nordstrom- AdS (anti-RN-(A)dS) black hole. We also study another phantom solution, which is called black plane \cite{R110}, and compare Quvedo's method  with our proposed  formulation. In  Section 4,  we study the Kerr Newman black holes \cite{R2} and show that the phase transitions  exactly occur on identical  points of the curvature singularities. In the final Section 5, we discuss our results.
\section{Specific heat and thermodynamic geometry singularities }
The first law of thermodynamics for RN black holes \cite{R22}, characterized by their mass $M$ and charge $Q$, can be  written as follows:
\begin{eqnarray}\label{103}
 dM=TdS+\Phi dQ
\end{eqnarray}
where, $\Phi$ is the potential difference between the horizon and infinity and $T$ is the Hawking temperature:
\begin{equation}\label{104}
T=(\frac{\partial M}{\partial S})_{Q}
\end{equation}
\begin{equation}\label{105}
\Phi =(\frac{\partial M}{\partial Q})_{S}
\end{equation}
Heat capacity at a constant potential is given by:
\begin{equation}\label{106}
C_{\Phi }=T(\frac{\partial S}{\partial T})_{\Phi }=\frac{1}{T(\frac{\partial T}{\partial S})}_{\Phi}
\end{equation}
Weinhold introduced a geometric formulation of  thermodynamics for the first time \cite{R21}. A few years later, Ruppeiner developed another geometric formulation for thermodynamics and statistical mechanics \cite{tun2}.
The Weinhold metric is defined as the second derivatives of the internal energy with respect to the entropy and other extensive parameters.
\begin{equation}\label{107}
g_{ij}^{W}=\frac{\partial ^{2}M(X^{K})}{\partial X^{i}\partial X^{J}};
\ \ \ X^{i}=\ (S,N^{a})
\end{equation}
where, $S$ is entropy and $N^{a}s$ determines all other extensive variables of the system.
The Ruppeiner metric is defined as the second derivatives of the entropy of the system with respect to the internal energy and other extensive variables, and is given by:
\begin{equation}\label{108}
g_{ij}^{R}=\frac{\partial^{2}S(X^{K})}{\partial X^{i}\partial X^{J}} ; \ \ \ X^{i}=(U,N^{a})
\end{equation}
where, $U$ is internal energy and $N^{a}s$ determines all other extensive variables of the system.
The line elements in the Weinhold and Ruppeiner geometries \cite{R33} are conformally related:
\begin{equation}\label{109}
dS_{R}^{2}=\frac{dS_{W}^{2}}{T}.
\end{equation}
The Ruppeiner metric of a black hole can be obtained by the following relations:
\begin{equation}\label{110}
g_{SS}=\frac{1}{T}(\frac{\partial ^{2}M}{\partial S^{2}})=\frac{1}{T}\frac{\partial }{\partial S}(\frac{\partial M}{\partial S})_{Q}=\frac{1}{T}(\frac{\partial T}{\partial S})_{Q}
\end{equation}
\begin{eqnarray}\label{111}
    &&g_{SQ}=\frac{1}{T}(\frac{\partial ^{2}M}{\partial Q\partial S})=\frac{1}{T}\frac{\partial }{\partial Q}
(\frac{\partial M}{\partial S})_{Q}\\
   \nonumber &&=\frac{1}{T}(\frac{\partial T}{\partial Q})_{S}=\frac{1}{T}(\frac{\partial \Phi }{\partial S})_{Q}=g_{QS}
     \end{eqnarray}
\begin{equation}\label{112}
g_{QQ}=\frac{1}{T}(\frac{\partial^{2}M}{\partial Q^{2}})=\frac{1}{T}\frac{\partial }{\partial Q}(\frac{\partial M}{\partial Q})_{S}=\frac{1}{T}(\frac{\partial \Phi }{\partial Q})_{S}
\end{equation}
where, $M$, $S$ and $Q$ are mass, entropy, and charge of the black hole, respectively.
In this Section we investigate the phase transition points of the heat capacity $C_{\Phi} $ at a constant electric potential and show that they  exactly correspond to the singularities of the scalar curvature $R(S,Q)$.
Using the first law, the following Maxwell relation can be obtained:
\begin{equation}\label{113}
(\frac{\partial T}{\partial Q})_{S}=(\frac{\partial \Phi }{\partial S})_{Q}
\end{equation}
We define $\overline{M}$ as a new conjugate potential of $M(S,Q)$ in order to determine another useful Maxwell equation. $\overline{M}$ is related to $M(S,Q)$ by the following Legendre transformation

\begin{equation}\label{114}
\overline{M}(S,\Phi )=M(S,Q)-\Phi Q\
\end{equation}
For this new function, the first law of thermodynamics will be:
\begin{equation}\label{115}
d\overline{M}=TdS-Qd\Phi \
\end{equation}
As a result, we can obtain another Maxwell relation as follows:
\begin{equation}\label{116}
(\frac{\partial T}{\partial \Phi })_{S}=-(\frac{\partial Q}{\partial S})_{\Phi }\
\end{equation}
Moreover, the metric elements for this conjugate potential are defined as in the following equations:
\beq
{{\overline{g}}_{SS}}=\frac{1}{T}\left( \frac{{{\partial }^{2}}\overline{M}}{{{\partial }^{2}}S} \right)=\frac{1}{T}\frac{\partial }{\partial S}{{(\frac{\partial \overline{M}}{\partial S})}_{\Phi }}=\frac{1}{T}{{(\frac{\partial T}{\partial S})}_{\Phi }}
\eeq
\bea
&&{{\overline{g}}_{S\Phi }}=\frac{1}{T}(\frac{{{\partial }^{2}}\overline{M}}{\partial \Phi \partial S})=\frac{1}{T}\frac{\partial }{\partial \Phi }{{(\frac{\partial \overline{M}}{\partial S})}_{\Phi }}=\frac{1}{T}{{(\frac{\partial T}{\partial \Phi })}_{S}}\\
\nonumber&&=-\frac{1}{T}{{(\frac{\partial Q}{\partial S})}_{\Phi }}={{\overline{g}}_{\Phi S}}
\eea
\beq
{{\overline{g}}_{\Phi \Phi }}=\frac{1}{T}(\frac{{{\partial }^{2}}\overline{M}}{\partial {{\Phi }^{2}}})=\frac{1}{T}\frac{\partial }{\partial \Phi }{{(\frac{\partial \overline{M}}{\partial \Phi })}_{S}}=-\frac{1}{T}{{(\frac{\partial Q}{\partial \Phi })}_{S}}
\eeq
The last part of the metric elements in (\ref{110}, \ref{111}, \ref{112}) are written by using Maxwell equations (\ref{113}) and (\ref{116}).
For calculating the scalar curvature for two dimensions associated with the Ruppeiner metric, we can use the following relation:
\begin{equation}\label{117}
R=\frac{\left| \begin{array}{ccc}
 \noalign{\medskip}  g_{SS } & g_{QQ  } & g_{SQ }  \\
\noalign{\medskip}   g_{SS,S } & g_{QQ ,S } & g_{SQ ,S }  \\
 \noalign{\medskip}  g_{SS ,Q} & g_{QQ ,Q } & g_{SQ ,Q }   \end{array}\right|}{-2\left| \begin{array}{cc}
 \noalign{\medskip}  g_{SS } & g_{SQ }  \\
\noalign{\medskip}   g_{SQ } & g_{QQ }  \end{array} \right|^{2}}
\end{equation}
The inverse of the heat capacity $C_{\Phi}$  can be written as:
\begin{equation}\label{118}
C_{\Phi }^{-1}=\frac{1}{T}(\frac{\partial T}{\partial S})_{\Phi }=-\frac{1}{T}(\frac{\partial T}{\partial \Phi })_{S}(\frac{\partial \Phi }{\partial S})_{T}
\end{equation}
We will prove that the square root of the denominator of $R(S,Q)$ is proportional to the inverse of $ C_{\Phi}$:
\begin{equation}\label{1r1}
(g_{SS}g_{QQ}-(g_{SQ})^{2})\propto -\frac{1}{T}(\frac{\partial T}{\partial \Phi })_{S}(\frac{\partial \Phi }{\partial S})_{T}
\end{equation}
The above equation means that the phase transition points correspond to the singularities of $R(S,Q)$.
Substituting the metric elements (\ref{110}, \ref{111}, \ref{112})  in the left hand side of (\ref{1r1}) and using (\ref{113}) yields
\begin{equation}\label{117}
-\frac{1}{T^{2}}(\frac{\partial T}{\partial Q})_{S}\left[
(\frac{\partial Q}{\partial S})_{T}(\frac{\partial \Phi }{\partial Q})_{S}+(\frac{\partial \Phi }{\partial S})_{Q}\right]
\end{equation}
Equation (\ref{117}) can be simplified by using the following equation:
\begin{equation}\label{118}
d\Phi =(\frac{\partial \Phi }{\partial S})_{Q}dS+(\frac{\partial \Phi }{\partial Q})_{S}dQ
\end{equation}
or
\begin{equation}\label{119}
(\frac{\partial \Phi }{\partial S})_{T}=(\frac{\partial \Phi }{\partial S})_{Q}+(\frac{\partial \Phi }{\partial Q})_{S}(\frac{\partial Q}{\partial S})_{T}
\end{equation}
Considering $T$ as a function of $Q$ and $S$, we have:
\begin{equation}\label{120}
dT=(\frac{\partial T}{\partial Q})_{S}dQ+(\frac{\partial T}{\partial S})_{Q}dS
\end{equation}
So, we get another useful equation:
\begin{equation}\label{121}
(\frac{\partial T}{\partial Q})_{S}=\frac{(\frac{\partial T}{\partial \Phi })_{S}}{(\frac{\partial Q}{\partial \Phi })_{S}}
\end{equation}
By substituting Eqs. (\ref{119}) and (\ref{121}) in (\ref{117}), the left hand side of (\ref{1r1}) can be simplified as:
\begin{equation}\label{122}
-\frac{1}{T^{2}}(\frac{(\frac{\partial T}{\partial \Phi })_{S}}
{(\frac{\partial Q}{\partial \Phi })_{S}})(\frac{\partial \Phi }{\partial S})_{T}=\frac{(\frac{\partial \Phi }{\partial Q})_{S}}{T}C_{\Phi }^{-1}
\end{equation}
We conclude that for a finite value of $(\partial \Phi /\partial Q)_{S}$  and $T\ne 0$, we have:
\begin{equation}\label{123}
\left| \begin{array}{cc}\noalign{\medskip}g_{SS } & g_{SQ }  \\
\noalign{\medskip}g_{SQ } & g_{QQ } \\
\end{array} \right|= \frac{(\frac{\partial \Phi }{\partial Q})_{S}}{T}C_\Phi ^{-1}
\end{equation}

\noindent This means that the singularities of $R(S,Q)$ correspond exactly to the phase transition points. A similar calculation for conjugate potential yields:
\beq\label{317}
\left| \begin{array}{cc}
   \noalign{\medskip}{{\overline{g}}_{SS}} & {{\overline{g}}_{S\Phi }}  \\
  \noalign{\medskip} {{\overline{g}}_{S\Phi }} & {{\overline{g}}_{\Phi \Phi }}  \\
\end{array} \right|=-\frac{{{(\frac{\partial Q}{\partial \Phi })}_{S}}}{T}C_{Q}^{-1}
\eeq
This means that the singularities of $\overline{R}(S,\Phi)$ are the same as transition points of $C_{Q}$ which identifies the Davies curve \cite{R27}. Moreover, the curvature singularity of the free-energy metric is also located at the Davies curve of $C_{Q}$.

 The Helmholtz free-energy is related to $M(S,Q)$ by the following Legendre transformation

\begin{equation}\label{314}
\overline{\overline{M}}(Q,T)=M(S,Q)-T S
\end{equation}
For this new function, the first law of thermodynamics can be expressed as
\begin{equation}\label{315}
d\overline{\overline{M}}(Q,T)=-SdT+\Phi dQ \
\end{equation}

The metric elements for the Helmholtz free-energy can be defined by the following equations:
\beq
{{\overline{\overline{g}}}_{TT}}=\frac{1}{T}\left( \frac{{{\partial }^{2}}\overline{\overline{M}}}{\partial {{T}^{2}}} \right)=-\frac{1}{T}{{\left( \frac{\partial S}{\partial T} \right)}_{Q}}
\eeq
\bea
&&{{\overline{\overline{g}}}_{TQ}}=\frac{1}{T}\left( \frac{{{\partial }^{2}}\overline{\overline{M}}}{\partial T\partial Q} \right)=-\frac{1}{T}{{\left( \frac{\partial S}{\partial Q} \right)}_{S}}\\
\nonumber &&=\frac{1}{T}{{\left( \frac{\partial \Phi }{\partial T} \right)}_{Q}}={{\overline{\overline{g}}}_{QT}}
\eea
\beq
{{\overline{\overline{g}}}_{TT}}=\frac{1}{T}\left( \frac{{{\partial }^{2}}\overline{\overline{M}}}{\partial {{Q}^{2}}} \right)=\frac{1}{T}{{\left( \frac{\partial \Phi }{\partial Q} \right)}_{T}}
\eeq
  We can also write the metric elements of Helmholtz free-energy in the same coordinates of conjugate potential $\overline{M}(S,\Phi)$ by using a transformation matrix. This matrix changes coordinates from $(T,Q)$ to $(S,\Phi)$. The transformation matrix can be written in the following form:
\begin{equation}
N=\left( \begin{array}{cc}
 \noalign{\medskip}  {{\left( \frac{\partial T}{\partial S} \right)}_{\Phi }} & {{\left( \frac{\partial T}{\partial \Phi } \right)}_{S}}  \\
   \noalign{\medskip} {{\left( \frac{\partial Q}{\partial S} \right)}_{\Phi }} & {{\left( \frac{\partial Q}{\partial \Phi } \right)}_{S}}  \\
\end{array} \right)
\end{equation}
  Using the transformation matrix, we can show that the metric elements of the  $\overline{\overline{M}}(T,Q)$ in new coordinate $(S,\Phi)$ is the same as the metric elements of the $\overline{M}(S,\Phi)$.
\begin{equation}
\overline{g}_{ij}'=N_{ik}^{T} \,\ {{\overline{\overline{g}}}_{kl}} \,\ {{N}_{lj}}
\end{equation}
where $N^{T}$ is transpose of $N$. Therefore, the singularity points of scalar curvature for both Helmholtz free-energy function and conjugate potential occur exactly at the same phase transition points of ${{C}_{Q }}$ (\ref{317}). In other words, the line element of free-energy $\overline{\overline{M}}(Q,T)$ is associated with the line element of the conjugate potential $\overline{M}(S,\Phi)$ \cite{R33}.
 \begin{equation}
 d{{s}^{2}}(\overline{\overline{M}})=-d{{s}^{2}}(\overline{M})
 \end{equation}
  Furthermore, we are able to define the metric element of $\overline{\overline{M}}$ in terms of the metric elements of $\overline{M}$ by below conformal transformation.
 \begin{equation}
 {{\overline{\overline{g}}}_{ij}}=-{{\overline{g}}_{ij}}=-\frac{1}{T} \left( \frac{{{\partial }^{2}}\overline{M}}{\partial {{X}^{i}}\partial {{X}^{j}}} \right) \ ;\ {{X}^{i}}=(S,\Phi )
 \end{equation}
The above calculations confirm our (trivial) expectation  that curvature is independent of any specific coordinate choice of the thermodynamic quantities.
We can also prove  that for black holes with three parameters the phase transition points of  heat capacity ${{C}_{\Phi ,\Omega }}$ at a constant electric potential and angular velocity correspond exactly to the singularities of the scalar curvature $R(S,Q,J)$. The scalar curvature is proportional to the inverse of the square determinant of the metric.
\begin{equation}\label{q2}
R(S,Q,J)\propto {{\left| \begin{array}{ccc}
 \noalign{\medskip}  {{g}_{SS}} & {{g}_{SQ}} & {{g}_{SJ}}  \\
 \noalign{\medskip}  {{g}_{QS}} & {{g}_{QQ}} & {{g}_{QJ}}  \\
 \noalign{\medskip}  {{g}_{JS}} & {{g}_{JQ}} & {{g}_{JJ}}  \\
\end{array} \right|}^{-2}}
\end{equation}

\noindent After a long but straightforward calculation, the inverse of the heat capacity ${{C}_{\Phi ,\Omega }}$ can be written as follows (see Appendix).

\bea\label{q3}\nonumber
&&\left[{{\left( \frac{\partial \Omega }{\partial J} \right)}_{S,Q}}{{\left( \frac{\partial \phi }{\partial Q} \right)}_{S,J}}-{{\left( \frac{\partial \Omega }{\partial Q} \right)}_{S,J}}{{\left( \frac{\partial \phi }{\partial J} \right)}_{S,Q}}\right]\\
\times &&\frac{{{({{C}_{\Phi ,\Omega }})}^{-1}}}{T^{2}}
=\left| \begin{array}{ccc}
 \noalign{\medskip}  {{g}_{SS}} & {{g}_{SQ}} & {{g}_{SJ}}  \\
 \noalign{\medskip}  {{g}_{QS}} & {{g}_{QQ}} & {{g}_{QJ}}  \\
 \noalign{\medskip}  {{g}_{JS}} & {{g}_{JQ}} & {{g}_{JJ}}  \\
\end{array} \right|
\eea

\noindent Therefore, we conclude that the singularities of $R(S,Q,J)$  correspond to the phase transition points of $C_{\Phi,\Omega}$ and that also those  of $\overline{R}(S,\Phi,\Omega)$  correspond to the phase transitions of $C_{Q,J}$.
\bea \label{q333}\nonumber
&&\left[ {{\left( \frac{\partial Q}{\partial \Phi } \right)}_{S,\Omega }}{{\left( \frac{\partial J}{\partial \Omega } \right)}_{S,\Phi }}-{{\left( \frac{\partial J}{\partial \Phi } \right)}_{S,\Omega }}{{\left( \frac{\partial Q}{\partial \Omega } \right)}_{S,\Phi }} \right]\\
\times&&\frac{{{({{C}_{Q,J}})}^{-1}}}{{{T}^{2}}}=\left| \begin{array}{ccc}
 \noalign{\medskip}  {{\overline{g}}_{SS}} & {{\overline{g}}_{S\Phi }} & {{\overline{g}}_{S\Omega }}  \\
 \noalign{\medskip}  {{\overline{g}}_{\Phi S}} & {{\overline{g}}_{\Phi \Phi }} & {{\overline{g}}_{\Phi \Omega }}  \\
 \noalign{\medskip}  {{\overline{g}}_{\Omega S }} & {{\overline{g}}_{ \Omega \Phi}} & {{\overline{g}}_{\Omega \Omega }}  \\
\end{array} \right|
\eea
We may generalize the above mentioned relations to a large class of black holes with an arbitrary number of parameters. The first law of thermodynamics for black holes with $(n+1)$ parameters can be written as follows:
\bea\label{r1}
dM=TdS+\sum\limits_{i=1}^{n}{{{\Phi }_{i}}{{Q}_{i}}}
\eea
It is clear that the energy $M$ is a function of $n+1$ extensive variables $(S,Q_{i})$. In addition, we can consider $(n+1)$ pairs of intensive/extensive variables $(T,S)$ and $(\Phi_{i},Q_{i})$. The Ruppenier metric for black holes with $(n+1)$ extensive variables can also be written by the following relation.
\beq\label{r2}
g_{ij}^{R}=\frac{1}{T}\left( \frac{{{\partial }^{2}}M}{\partial {{X}^{i}}\partial {{X}^{j}}} \right);{{X}^{i}}=(S,{{Q}_{1}},{{Q}_{2}},...,{{Q}_{n}})
\eeq
In general, we contend that the transition points of heat capacity $C_{\Phi_{1},...,\Phi_{n}}$ are the same as singularity points of the scalar curvature $R(S,Q_{1},...,Q_{n})$. The scalar curvature $R(S,Q_{1},...,Q_{n})$ is proportional to the inverse of the square determinant of metric.
\beq\label{r3}
R(S,{{Q}_{1}},...,{{Q}_{n}})\propto \left| \begin{array}{cccc}
 \noalign{\medskip}   {{g}_{S  S{}}} & {{g}_{S{{Q}_{1}}}} & ... & {{g}_{S{{Q}_{n}}}}  \\
  \noalign{\medskip}  {{g}_{{{Q}_{1}}S}} & {{g}_{{{Q}_{1}}{{Q}_{1}}}} & ... & {{g}_{{{Q}_{1}}{{Q}_{n}}}}  \\
  \noalign{\medskip}  : & : & : & :  \\
  \noalign{\medskip}  {{g}_{{{Q}_{n}}S}} & {{g}_{{{Q}_{n}}{{Q}_{1}}}} & ... & {{g}_{{{Q}_{n}}{{Q}_{n}}}}  \\
\end{array} \right|^{-2}
\eeq
Using equation (\ref{r3}), It is expected  that the inverse of the heat capacity $C_{\Phi_{1},...,\Phi_{n}}$ can be obtained by below relation.
\nonumber\bea\nonumber
\left[ \frac{{{({{C}_{ \Phi_{1},\Phi_{2},...,\Phi_{n} }})}^{-1}}}{{{T}^{n}}}\left( \frac{\partial (\Phi_{1},\Phi_{2},... ,\Phi_{n} )}{\partial {{(Q_{1},Q_{2},...,Q_{n})}_{S}}} \right) \right ]
\eea
\bea\label{r555}
  =\left| \begin{array}{cccc}
 \noalign{\medskip}   {{g}_{S S{}}} & {{g}_{S{{Q}_{1}}}} & ... & {{g}_{S{{Q}_{n}}}}  \\
  \noalign{\medskip}  {{g}_{{{Q}_{1}}S}} & {{g}_{{{Q}_{1}}{{Q}_{1}}}} & ... & {{g}_{{{Q}_{1}}{{Q}_{n}}}}  \\
  \noalign{\medskip}  : & : & : & :  \\
  \noalign{\medskip}  {{g}_{{{Q}_{n}}S}} & {{g}_{{{Q}_{n}}{{Q}_{1}}}} & ... & {{g}_{{{Q}_{n}}{{Q}_{n}}}}  \\
\end{array} \right|
\eea
where
\beq
\frac{\partial ({{\Phi }_{1}},{{\Phi }_{2}},...,{{\Phi }_{n}})}{\partial {{({{Q}_{1}},{{Q}_{2}},...,{{Q}_{n}})}_{S}}}=\left| \begin{array}{cccc}
 \noalign{\medskip}  \left( \frac{\partial {{\Phi }_{1}}}{\partial {{Q}_{1}}} \right) & \left( \frac{\partial {{\Phi }_{1}}}{\partial {{Q}_{2}}} \right) & ... & \left( \frac{\partial {{\Phi }_{1}}}{\partial {{Q}_{n}}} \right)  \\
 \noalign{\medskip}  \left( \frac{\partial {{\Phi }_{2}}}{\partial {{Q}_{1}}} \right) & \left( \frac{\partial {{\Phi }_{2}}}{\partial {{Q}_{2}}} \right) & ... & \left( \frac{\partial {{\Phi }_{2}}}{\partial {{Q}_{n}}} \right)  \\
 \noalign{\medskip}  : & : & : & :  \\
 \noalign{\medskip}  \left( \frac{\partial {{\Phi }_{n}}}{\partial {{Q}_{1}}} \right) & \left( \frac{\partial {{\Phi }_{n}}}{\partial {{Q}_{2}}} \right) & ... & \left( \frac{\partial {{\Phi }_{n}}}{\partial {{Q}_{n}}} \right)  \\
\end{array} \right|
\eeq

\noindent Thus, the singularities of $R(S,Q_{1},...,Q_{n})$  correspond to the phase transition points of $C_{\Phi_{1},...,\Phi_{n}}$. Although we have proved equations (\ref{q3}) and (\ref{q333}), we do not know a rigorous proof for equation (\ref{r555}) at this time.  A general proof for equation (\ref{r555}) remains as an open problem. We hope to solve this problem in the near future. On the other hand, the conjugate potential $\overline{M}(S,\Phi_{1},...,\Phi_{n} )$ can be obtained from  $M(S,Q_{1},...,Q_{n})$ by the following Legendre transformation:
\begin{equation}\label{r5}
\overline{M}(S,\Phi_{1} ,...,\Phi_{n} )=M(S,Q_{1},...,Q_{n})-\sum\limits_{i=1}^{n}{{{\Phi }_{i}}{{Q}_{i}}}
\end{equation}
by defining the metric elements for the conjugate  potential $\overline{M}(S,\Phi_{1},...,\Phi_{n} )$ as:
\beq
{{\overline{g}}_{ij}}=\frac{1}{T}\left( \frac{{{\partial }^{2}}\overline{M}}{\partial {{X}^{i}}\partial {{X}^{j}}} \right);{{X}^{i}}=(S,{{\Phi }_{1}},{{\Phi }_{2}},...,{{\Phi }_{n}})
\eeq
we can assert that the singularity points of \\
 $\overline{R}(S,\Phi_{1},\Phi_{2},...,\Phi_{n})$ correspond to the phase transitions of $C_{Q_{1},Q_{2},...,Q_{n}}$.
 \bea\nonumber
\frac{{{({{C}_{ Q_{1},Q_{2},...,Q_{n} }})}^{-1}}}{{{T}^{n}}}\left( \frac{\partial (Q_{1},Q_{2},...,Q_{n} )}{\partial {{(\Phi_{1},\Phi_{2},...,\Phi_{n})}_{S}}} \right)
\eea
\bea
 =\left| \begin{array}{cccc}
 \noalign{\medskip}   {{\overline{g}}_{S S }} & {{\overline{g}}_{S{{\Phi}_{1}}}} & ... & {{\overline{g}}_{S{{\Phi}_{n}}}}  \\
  \noalign{\medskip}  {{\overline{g}}_{{{\Phi}_{1}}S}} & {{\overline{g}}_{{{\Phi}_{1}}{{\Phi}_{1}}}} & ... & {{\overline{g}}_{{{\Phi}_{1}}{{\Phi}_{n}}}}  \\
  \noalign{\medskip}  : & : & : & :  \\
  \noalign{\medskip}  {{\overline{g}}_{{{\Phi}_{n}}S}} & {{\overline{g}}_{{{\Phi}_{n}}{{\Phi}_{1}}}} & ... & {{\overline{g}}_{{{\Phi}_{n}}{{\Phi}_{n}}}}  \\
\end{array} \right|
\eea
\begin{table*}
\begin{tabular*}{\textwidth}{@{\extracolsep{\fill}}lrrl@{}}
\hline
$kerr$ & $RN$ \\
\hline
$M(S,J)=\sqrt{\frac{S}{4\pi }+\frac{J^{2}\pi}{S}}$ & $M(S,Q)=\frac{\sqrt{S\pi}}{2}(\frac{1}{\pi}+\frac{Q^{2}}{S})$  \\
$T(S,J)=\frac{{{S}^{2}}-4{{J}^{2}}{{\pi }^{2}}}{4{{S}^{3/2}}\sqrt{\pi {{S}^{2}}+4{{J}^{2}}{{\pi }^{3}}}}$ & $T(S,Q)=\frac{S-{{Q}^{2}}\pi }{4{{S}^{3/2}}\sqrt{\pi }}$\\
$R(S,J)=-\frac{S({{S}^{2}}+12{{\pi }^{2}}{{J}^{2}})}{({{S}^{2}}+4{{\pi }^{2}}{{J}^{2}})(4{{\pi }^{2}}{{J}^{2}}-{{S}^{2}})}$ & $R(S,Q)=0$ \\
$C_{\Omega}(S,J)=2\frac{\left( -{{S}^{2}}+4{{J}^{2}}{{\pi }^{2}} \right){{S}^{3}}}{{{\left( {{S}^{2}}+4{{J}^{2}}{{\pi }^{2}} \right)}^{2}}}$ & $C_{\Phi}(S,Q)=-2S$\\
\hline
$\overline{M}(S,\Omega)=\sqrt{\frac{S}{4\pi }-(\frac{S\Omega}{2\pi })^{2}}$ & $\overline{M}(S,\Phi)=\sqrt{\frac{S}{4\pi}}(\Phi ^{2}-1)$ \\
 $T(S,\Omega)=\frac{-2S{{\Omega }^{2}}+\pi }{4\pi \sqrt{S(-S{{\Omega }^{2}}+\pi })}$&  $T(S,\Phi)=\frac{1-{{\Phi }^{2}}}{4\sqrt{\pi S}}$\\
$\overline{R}(S,\Omega)=4\frac{\left( -S{{\Omega }^{2}}+\pi  \right){{\pi }^{2}}\left( {{\pi }^{2}}-2S{{\Omega }^{2}}\pi -8{{S}^{2}}{{\Omega }^{4}} \right)}{{{\left( {{\pi }^{2}}-8S{{\Omega }^{2}}\pi +4{{S}^{2}}{{\Omega }^{4}} \right)}^{2}}\left( -2S{{\Omega }^{2}}+\pi  \right)S}$ & $\overline{R}(S,\Phi )=-\frac{-1+{{\Phi }^{2}}}{{{\left( -1+3{{\Phi }^{2}} \right)}^{2}}S}$
 \\
${{C}_{J}}(S,\Omega )=-2\frac{\pi S\left( -2S{{\Omega }^{2}}+\pi  \right)}{4{{S}^{2}}{{\Omega }^{4}}-8S{{\Omega }^{2}}\pi +{{\pi }^{2}}}$ & ${{C}_{Q}}(S,\Phi )=-2\frac{S\left( {{\Phi }^{2}}-1 \right)}{-1+3{{\Phi }^{2}}}$\\

\hline
\end{tabular*}
\caption{  Thermodynamic variables and scalar curvature functions for $Kerr$ and $RN$ ($Reissner-Nordstrom$) black holes.}
\end{table*}\label{aa}
\begin{table*}
\begin{tabular*}{\textwidth}{@{\extracolsep{\fill}}lrrl@{}}
\hline
$BTZ$&$EMGB$\\
\hline
$M(S,J)=\frac{{{S}^{2}}}{16{{\pi }^{2}}{{l}^{2}}}+\frac{4{{\pi }^{2}}{{J}^{2}}}{{{S}^{2}}}$ & $M(S,Q)=\pi \alpha +\frac{\pi {{Q}^{2}}}{6\sqrt[3]{{{S}^{2}}}}+{{\pi }^{2}}\sqrt[3]{{{S}^{2}}}-\frac{\pi \Lambda \sqrt[3]{{{S}^{4}}}}{12}$\\
$T(S,J)=\frac{{{S}^{4}}-64{{\pi }^{4}}{{J}^{2}}{{L}^{2}}}{8{{\pi }^{2}}{{L}^{2}}{{S}^{3}}}$&$T(S,Q)=\frac{\pi \left( -{{Q}^{2}}+3{{S}^{4/3}}-\Lambda {{S}^{2}} \right)}{9{{S}^{5/3}}}$\\
$R(S,J)=0$&$R(S,Q)=-\frac{\Lambda \left( -6{{S}^{10/3}}{{Q}^{2}}+{{\Lambda }^{2}}{{S}^{6}}-9{{S}^{14/3}}+3{{Q}^{4}}{{S}^{2}}-4{{Q}^{2}}\Lambda {{S}^{4}}+8{{S}^{16/3}}\Lambda  \right)}{S{{\left( -{{Q}^{2}}+3{{S}^{4/3}}+\Lambda {{S}^{2}} \right)}^{2}}\left( -{{Q}^{2}}+3{{S}^{4/3}}-\Lambda {{S}^{2}} \right)}$\\
$C_{\Omega}(S,J)= S$&$C_{\Phi}(S,Q)=-3\frac{\left( -{{Q}^{2}}+3{{S}^{4/3}}-\Lambda {{S}^{2}} \right)S}{-{{Q}^{2}}+3{{S}^{4/3}}+\Lambda {{S}^{2}}}$\\
\hline

$\overline{M}(S,\Omega)=\frac{S^{2}}{16\pi^{2}l^{2}}-\frac{\Omega^{2}S^{2}}{16\pi^{2}}$ &$\overline{M}(S,\Phi)=\pi\alpha -\frac{3\Phi \sqrt[3]{S^{2}}}{\pi }+\frac{\pi \sqrt[3]{S^{2}}}{12}-\frac{\pi \Lambda \sqrt[3]{S^{4}}}{12}$\\
$T(S,\Omega)=-\,{\frac {S \left( -1+{\Omega}^{2}{L}^{2} \right) }{8{\pi }^{2}{L}^
{2}}}$
&$T(S,\Phi )=-\frac{9{{\Phi }^{2}}-3{{\pi }^{2}}+{{\pi }^{2}}\Lambda {{S}^{2/3}}}{9\pi \sqrt[3]{S}}$\\
$\overline{R}(S,\Omega )=2\frac{-1+{{\Omega }^{2}}{{L}^{2}}}{{{\left( 1+3{{\Omega }^{2}}{{L}^{2}} \right)}^{2}}S}$&$\overline{R}(S,\Phi )=-\frac{A(S,\Phi )}{{{S}^{5/3}}{{\left( -45{{\Phi }^{2}}+3{{\pi }^{2}}+{{\pi }^{2}}\Lambda {{S}^{2/3}} \right)}^{2}}\left( 9{{\Phi }^{2}}-3{{\pi }^{2}}+{{\pi }^{2}}\Lambda {{S}^{2/3}} \right)}$\\
${{C}_{J}}(S,\Omega )=-\frac{S\left( -1+{{\Omega }^{2}}{{L}^{2}} \right)}{1+3{{\Omega }^{2}}{{L}^{2}}}$&${{C}_{Q}}(S,\Phi )=3S\frac{\left( 9{{\Phi }^{2}}-3{{\pi }^{2}}+{{\pi }^{2}}\Lambda {{S}^{2/3}} \right)}{-45{{\Phi }^{2}}+3{{\pi }^{2}}+{{\pi }^{2}}\Lambda {{S}^{2/3}}}$\\
\hline
\end{tabular*}
\caption{Thermodynamic variables and scalar curvature functions for $BTZ$ ($Banados-Teitelboim-Zanelli$)and $EMGB$ ($Einstein-Maxwell-Gauss-Bonnet$) black holes. We use the following notations, $\Lambda$= cosmological constant, $k$ = Chern-Simons coupling constant and $\alpha$ = Gauss-Bonnet coupling constant.}
\end{table*}\label{bb}
\begin{table*}
\begin{tabular*}{\textwidth}{@{\extracolsep{\fill}}lrrl@{}}
\hline
$EGB$&$EYMGB$\\
\hline
$M(S,Q)=\sqrt[3]{S^{2}}+\frac{Q^{2}}{3\sqrt[3]{S^{2}}}$ & $M(S,Q)=\sqrt[3]{S^{2}}-\frac{2Q^{2}\ln (S)}{3}$\\
$T(S,Q)=\frac{2(3{{S}^{4/3}}-{{Q}^{2}})}{9{{S}^{5/3}}}$&$T(S,Q)=\frac{2(S-{{Q}^{2}}\sqrt[3]{S})}{3{{S}^{4/3}}}$\\
$R(S,Q)=0$&$R(S,Q)=-\frac{B(S,Q)}{6\left( -S+{{Q}^{2}}\sqrt[3]{S} \right){{\left( -\ln \left( S \right)S+3\ln \left( S \right){{Q}^{2}}\sqrt[3]{S}+6{{Q}^{2}}\sqrt[3]{S} \right)}^{3}}}$\\
$C_{\Phi}(S,Q)= -3S$&$C_{\Phi}(S,Q)=-\frac{3\ln \left( S \right)\left( -S+{{Q}^{2}}\sqrt[3]{S} \right)S}{-\ln \left( S \right)S+6{{Q}^{2}}\sqrt[3]{S}+3{{Q}^{2}}\ln \left( S \right)\sqrt[3]{S}}$\\
\hline

$\overline{M}(S,\Phi)=\sqrt[3]{S^{2}}(1-\frac{3}{4}\Phi^{2})$ &$\overline{M}(S,\Phi)=\sqrt[3]{S^{2}}+\frac{3\Phi^{2}}{8\ln(S)}$\\
$T(S,\Phi )=-\frac{-4+3{{\Phi }^{2}}}{6\sqrt[3]{S}}$
&$T(S,\Phi )=-\frac{-16{{\left( \ln \left( S \right) \right)}^{2}}S+9{{\Phi }^{2}}\sqrt[3]{S}}{24{{S}^{4/3}}{{\left( \ln \left( S \right) \right)}^{2}}}$\\
$\overline{R}(S,\Phi )=-8\frac{-4+3{{\Phi }^{2}}}{{{\left( -4+15{{\Phi }^{2}} \right)}^{2}}S}$&$\overline{R}(S,\Phi ) -\frac{D(S,\Phi )}{{{\left( \ln \left( S \right) \right)}^{2}}\left( -16{{\left( \ln \left( S \right) \right)}^{2}}S+9{{\Phi }^{2}}\sqrt[3]{S} \right){{S}^{\frac{20}{3}}}{{\left( -16{{\left( \ln \left( S \right) \right)}^{2}}{{S}^{2}}+27{{\Phi }^{2}}{{S}^{4/3}} \right)}^{3}}}$\\
${{C}_{Q}}(S,\Phi )=-3\frac{\left( -4+3{{\Phi }^{2}} \right)S}{-4+15{{\Phi }^{2}}}$&${{C}_{Q}}(S,\Phi )=-3\frac{S\left( -16{{\left( \ln \left( S \right) \right)}^{2}}S+9{{\Phi }^{2}}\sqrt[3]{S} \right)}{-16{{\left( \ln \left( S \right) \right)}^{2}}S+27{{\Phi }^{2}}\sqrt[3]{S}}$\\
\hline
\end{tabular*}
\caption{Thermodynamic variables and scalar curvature functions for $EGB$ ($Einstein-Gauss-Bonnet$) and $ EYMGB$ ( $Einstein-Yang-Mills-Gauss-Bonnet$) black holes. }
\end{table*}\label{bbb}

We have collected mass ($M$), the conjugate potential $(\overline{M})$,  capacities, scalar curvature functions, and  temperature for various black holes and listed them in Tables 1, 2 and 3.  It is clear from these Tables  that both the heat capacity and  the scalar curvature diverge at the same point.For {\bf Kerr}, {\bf EMGB}, and {\bf EYMGB} balck holes, a factor proportional to temperature appears  in the denominator of the Ricci scalar. In order to probe this argument, we start with  the following metric
 \begin{equation}
 {{\overline{g}}_{\gamma \beta }}=\frac{1}{T}\left( \frac{{{\partial }^{2}}\overline{M}}{\partial \gamma \partial \beta } \right)=\frac{1}{T}{{h}_{\gamma \beta }}
 \end{equation}
 to show that
  \begin{equation}
 \overline{g}_{\gamma \beta ,\beta }^{{}}={{\partial }_{\beta }}(\frac{1}{T}{{h}_{\gamma \beta }})=\frac{1}{{{T}^{2}}}\tilde{h}_{\gamma \beta ,\beta }^{{}}
 \end{equation}
 where
 \begin{equation}
 \tilde{h}_{\gamma \beta ,\beta }^{{}}=-h_{\gamma \beta }^{{}}{{\partial }_{\beta }}T+Th_{\gamma \beta ,\beta }^{{}}
 \end{equation}
By replacing these equations in the numerator of the scalar curvature, we have:
 \begin{equation}
\left| \begin{array}{ccc}
  \noalign{\medskip} \frac{1}{T}h_{\beta \beta }^{{}} & \frac{1}{T}h_{\gamma \gamma }^{{}} & \frac{1}{T}h_{\gamma \beta }^{{}}  \\
  \noalign{\medskip} \frac{1}{{{T}^{2}}}\tilde{h}_{\beta \beta ,\beta }^{{}} & \frac{1}{{{T}^{2}}}\tilde{h}_{\gamma \gamma ,\beta }^{{}} & \frac{1}{{{T}^{2}}}\tilde{h}_{\gamma \beta ,\beta }^{{}}  \\
  \noalign{\medskip} \frac{1}{{{T}^{2}}}\tilde{h}_{\beta \beta ,\gamma }^{{}} & \frac{1}{{{T}^{2}}}\tilde{h}_{\gamma \gamma ,\gamma }^{{}} & \frac{1}{{{T}^{2}}}\tilde{h}_{\gamma \beta ,\gamma }^{{}}  \\
\end{array} \right|\propto \frac{1}{{{T}^{5}}}
\end{equation}
Also for the denominator of the scalar curvature we have:
\begin{equation}
 {{\left| \begin{array}{cc}
 \noalign{\medskip}  \frac{1}{T}h_{\gamma \gamma }^{{}} & \frac{1}{T}h_{\gamma \beta }^{{}}  \\
 \noalign{\medskip}  \frac{1}{T}h_{\gamma \beta }^{{}} & \frac{1}{T}h_{\beta \beta }^{{}}  \\
\end{array} \right|}^{2}}\propto \frac{1}{{{T}^{4}}}
 \end{equation}
 And consequently:
 \begin{equation}
\overline{R}\propto \frac{1}{T}\
 \end{equation}
 On the other hand, for {\bf RN}, {\bf BTZ}, and {\bf EGB} black holes, two elements of the metric are equal to zero.
For RN and EGB black holes,  we have ${{\partial }_{S}}(\bar{g}_{S\Phi })={{\partial }_{\Phi }}(\bar{g}_{SS})=0$ and for BTZ, ${{\partial }_{S}}(\bar{g}_{S\Omega })={{\partial }_{\Omega }}(\bar{g}_{SS})=0$. Therefore, the numerator of the scalar curvature is proportional to $\frac{1}{{{T}^{3}}}$ and the scalar curvature is proportional to temperature ($\overline{R}\propto T$).
\section{Thermodynamic geometry of Phantom Reissner-Nordestrom-AdS and black plane}

Recently, a new solution of Enistein-anti-Maxwell theory with a cosmological constant,  called the anti-Reissner-Nordstrom-(A)de Sitter solution, has been investigated \cite{R1}. This new solution has led to the following thermodynamic expression for the mass of this black hole.
\bea
M=\frac{1}{2}{{(S/\pi )}^{3/2}}(\frac{\pi }{S}-\frac{\Lambda }{3}+\frac{\eta {{\pi }^{2}}{{Q}^{2}}}{{{S}^{2}}})\
\label{ss2}\eea
where, $\Lambda$ is the cosmological constant, which can behave as $\Lambda > 0$ (dS) or $\Lambda < 0$ (AdS). At $\eta$=1, we have a solution for Reissner- Nordstrom-A dS, while $\eta$=-1, due to the negative energy of the field of Spin 1, gives us a solution for anti-Reissner-Nordstrom-AdS (phantom). The Hawking temperature, $T$, the electric potential, $\Phi$, and $C_{Q}$ are defined as follows:
\bea
T=(\frac{\partial M}{\partial S})=\frac{-\pi S+\Lambda {{S}^{2}}+\eta {{\pi }^{2}}{{Q}^{2}}}{-4{{(\pi S)}^{3/2}}}\
\eea
\bea
\Phi =(\frac{\partial M}{\partial Q})=\frac{{{(S/\pi )}^{3/2}}\eta {{\pi }^{2}}Q}{{{S}^{2}}}\
\label{mm2}\eea
\bea
&&{{C}_{Q}}=T{{(\frac{\partial S}{\partial T})}_{Q}}=\frac{T}{{{(\frac{\partial T}{\partial S})}_{Q}}}\\
&&\nonumber=\frac{-2S(-\pi S+\Lambda {{S}^{2}}+\eta {{\pi }^{2}}{{Q}^{2}})}{(-\pi S-\Lambda {{S}^{2}}+3\eta {{\pi }^{2}}{{Q}^{2}})}\
\label{mm1}\eea
In \cite{quved}, geomethermodynamic approach is used to obtain phase transition points. However, this theory is not able to produce the correct phase transition points.
In summary, the geometrothermodynamics of black holes is considered as a $2n + 1$ dimensional thermodynamic phase space, $T$, with independent coordinates ${\Phi ,E^{a},I^{a}}$, $a=1... n$, where $\Phi$  represents the thermodynamic potential, and $E^{a}$ and  $I^{a}$ are the extensive and  intensive thermodynamic variables, respectively. If the space $T$ possesses a non-degenerated metric $G_{AB}(Z^{C})$, where $Z^{c}={\Phi, E^{a}, I^{a}}$; and  one form of Gibbs $\Theta =d\Phi -\delta _{ab}I^{a}E^{b}$ in which  $\delta _{ab}$is the kornecher delta, then:
\bea
&&G=(d\Phi -{{\delta }_{ab}}{{I}^{a}}d{{E}^{b}})+({{\delta }_{ab}}{{E}^{a}}{{I}^{b}})({{\eta }_{cd}}d{{E}^{c}}d{{I}^{d}})\\
 &&\nonumber{{\eta }_{cd}}=diag(-1,1,....,1)\
\eea
Gibbs's form is invariant under  Legendre transformations that is written as:
\bea
\{\Phi ,{{E}^{a}},{{I}^{a}}\}\to \{\tilde{\Phi },{{\tilde{E}}^{a}},{{\tilde{I}}^{a}}\}&\Phi =\tilde{\Phi }-{{\delta }_{ab}}{{\tilde{E}}^{a}}{{\tilde{I}}^{b}}\
\eea
where:
\bea
{{E}^{a}}=-{{\tilde{I}}^{a}}&{{I}^{a}}={{\tilde{E}}^{a}}\
\eea
On the other hand, if somebody considers a n-dimensional subspace $E$ such that $E\subset T$,  we will, therefore, obtain  $d\Phi =\delta _{ab}I^{a}dE^{b}$ which is called the first law of thermodynamics. In  space $E$, the Quevedo metric is given by:
\bea
{{g}^{Q}}=({{E}^{c}}\frac{\partial \Phi }{\partial {{E}^{c}}})({{\eta }_{ab}}{{\delta }^{bc}}\frac{{{\partial }^{2}}\Phi }{\partial {{E}^{c}}\partial {{E}^{d}}}d{{E}^{a}}d{{E}^{b}})\
\label{ss1}\eea
Using (\ref{ss1}) and (\ref{ss2}), the scalar curvature is obtained as follows:
\bea
&&R(S,Q)=\frac{A(S,Q)}{{{(S\pi +\Lambda {{S}^{2}}-3{{\pi }^{2}}\eta {{Q}^{2}})}^{2}}}\\
&&\nonumber \times \frac{1}{{{(-S\pi +\Lambda {{S}^{2}}-3{{\pi }^{2}}\eta {{Q}^{2}})}^{3}}}
\eea
where, the points $S_{1}=-(\frac{\pi}{2 \Lambda})(1+\sqrt{1+12 \eta \Lambda Q^{2}})$, \\$S_{2}=-(\frac{\pi}{2 \Lambda})(-1+\sqrt{1+12 \eta \Lambda Q^{2}})$ and $ S_{3}=-(\frac{\pi}{ 2 \Lambda})(1-\sqrt{1+12 \eta \Lambda Q^{2}})$ are singularities of $R(S,Q)$.
All the other points are negative or a complex value for the entropy, and have, thus,  been rejected. The points of phase transition of specific heat in equation (\ref{mm1}) are only $S_{1}$ and $S_{3}$ . However, the extra point, i.e. $S_{2}$,  does not correspond to a  phase transition. Thus, the geometrodynamic method is not able to provide the same result as does the analysis  by using the heat capacity of RN-AdS and anti-RN-AdS black holes.

Using a Legendre transformation, we can find a conjugate potential for $M(S,Q)$.
\bea
\overline{M}(S,\Phi )=M(S,Q)-\Phi Q\
\label{mm3}\eea
By solving $Q$ from (\ref{mm2}) and substituting it in (\ref{mm3}), we have:
\bea
\overline{M}(S,\Phi)=\frac{1}{2}{{(\frac{S}{\pi })}^{\frac{3}{2}}}(\frac{\pi }{S}-\frac{\Lambda }{3}+\frac{\pi {{\Phi }^{2}}}{\eta S})-\frac{{{\Phi }^{2}}{{S}^{2}}}{{{(S/\pi )}^{3/2}}\eta {{\pi }^{2}}}\
\eea
We define the following metric:
\beq
{{\overline{g}}_{ij}}=\frac{1}{T}\left( \frac{{{\partial }^{2}}\overline{M}}{\partial {{X}^{i}}\partial {{X}^{j}}} \right) ; {{X}^{i}}=(S,\Phi )
\eeq
 Therefore, we will obtain the scalar curvature, $R$, as a function of $S$ and $\Phi$. Finally, using the Relation (\ref{mm2}), we can rewrite $R$ as a function of $S$ and $Q$.
\bea
&&\bar{R}(S,Q)=\frac{C(S,Q)}{{{(-S\pi -\Lambda {{S}^{2}}+3{{\pi }^{2}}\eta {{Q}^{2}})}^{2}}}\\
&&\nonumber \times \frac{1}{(-S\pi +\Lambda {{S}^{2}}+{{\pi }^{2}}\eta {{Q}^{2}})}
\label{nn1}\eea
The roots of the first part of the denominator gives us  $S_{1}$ and $S_{3}$; i.e, the phase transition points. The second part of the denominator is only zero at $T=0$ or for extremal black holes. Therefore, the curvature diverges exactly at these points where  heat capacity diverges with no other additional roots.

\begin{figure}[tbp]
\centering
\fbox{\includegraphics[scale=.5]{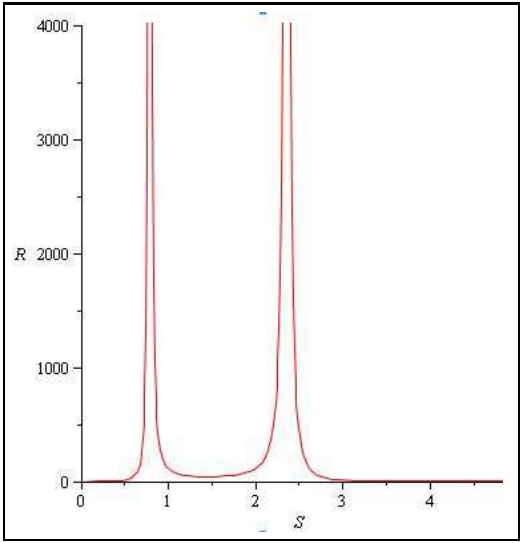} }
\caption{Graph of the scalar of curvature as a  function of  entropy, $S$, in the {\bf RN-AdS} case, for
an electric charge $Q = 0.25$ and a cosmological constant $\Lambda$  = -1.\label{fig1} }
\end{figure}

\begin{figure}[tbp]
\centering
\fbox{\includegraphics[scale=.5]{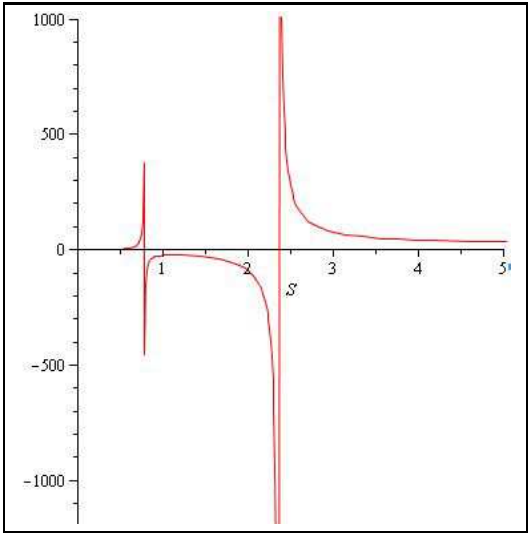} }
\caption{ Graph of the specific heat as a function of  entropy, S, in the {\bf RN-AdS} case, for an
electric charge $Q = 0.25$ and a cosmological constant $\Lambda $ =-1. \label{fig2} }
\end{figure}

For the {\bf RN-AdS} black hole, the scalar  curvature (\ref{nn1}) and the specific
heat (\ref{mm1}) are depicted in Figures \ref{fig1} and \ref{fig2}, respectively, as a function of  entropy and for a fixed value of  electric charge
$Q = 0.25$.

 The Ruppeiner curvature can also be used to
probe the microstructure of a thermodynamic system \cite{mirza2,R53}. The scalar curvature is positive in Figure \ref{fig1}; we, therefore, expect a fermion-like or short range repulsive behavior for the microstructure of {\bf RN-AdS} black holes.

\begin{figure}[tbp]
\centering
\fbox{\includegraphics[scale=.5]{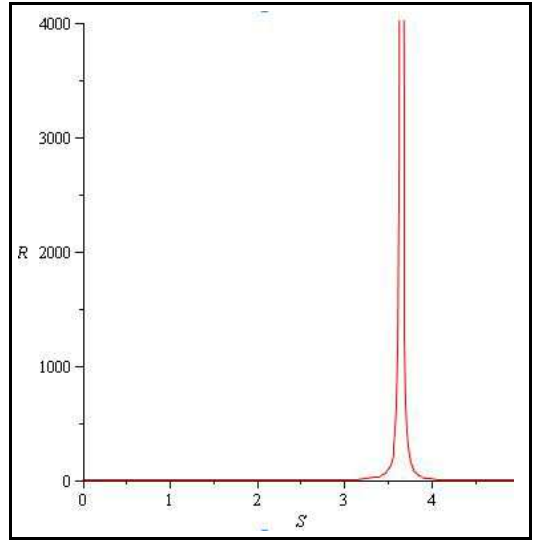} }
\caption{Graph of the scalar of curvature as a  function of  entropy, $S$, of the {\bf Phantom Reissner-Nordestrom-AdS} black hole for an
electric charge $Q = 0.25$ and a cosmological constant $\Lambda$  = -1. \label{fig3} }
\end{figure}

\begin{figure}[tbp]
\centering
\fbox{\includegraphics[scale=.5]{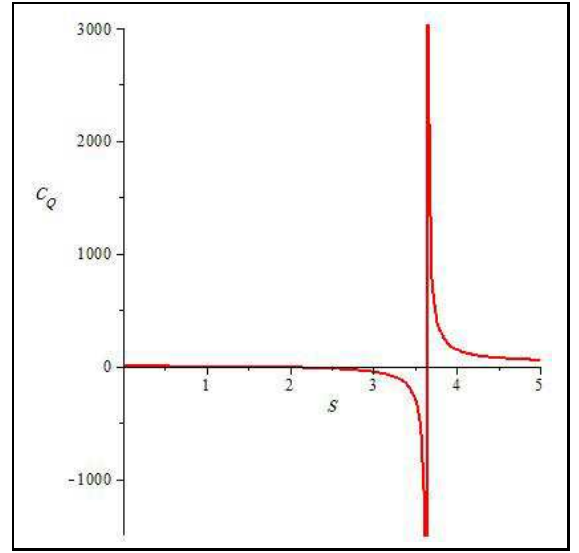} }
\caption{Graph of the specific heat as a  function of  entropy, $S$, of the {\bf Phantom Reissner-Nordestrom-AdS} black hole for an electric charge $Q = 0.25$ and a cosmological constant $\Lambda$  =-1. \label{fig4} }
\end{figure}

A change of sign
for heat capacity is usually associated with a drastic change in the stability properties of a
thermodynamic system; a negative heat capacity represents a region of instability whereas
the stable domain is characterized by a positive heat capacity. For RN-AdS black hole, the  unstable region (${{C}_{Q}} < 0$) is between  $S_{1}$ and $S_{3}$ while we expect stability for $S < {{S}_{1}}$ and $S > {{S}_{3}}$.
 The scalar curvature and the specific heat for {\bf Phantom Reissner-Nordestrom-AdS} are depicted in  Figures \ref{fig3} and \ref{fig4}, respectively.

The four dimensional black plane is another interesting thermodynamic system with two degrees of freedom \cite{R110}. The mass of the black plane is given by:
\bea
M(S,Q)=\frac{{{\alpha }^{2}}{{S}^{2}}+\eta {{\pi }^{2}}{{Q}^{2}}}{\pi \alpha \sqrt{2S}}\
\label{bb1}\eea
where, $\Lambda$ is the cosmological constant and ${{\alpha }^{2}}=\frac{-\Lambda }{3}$. When considering $\eta$=1, we have a solution for the normal black plane, while $\eta$=-1  gives us a solution for the phantom black plane. Also the temperature and the electric potential can be written in terms of the entropy and the electric charge as:
\bea
T=(\frac{\partial M}{\partial S})=\frac{3{{\alpha }^{2}}{{S}^{2}}-\eta {{\pi }^{2}}{{Q}^{2}}}{\pi \alpha {{(\sqrt{2S})}^{3}}}\
\eea
\bea
\Phi =(\frac{\partial M}{\partial Q})=\frac{2\pi Q}{\alpha \sqrt{2S}}\
\label{kk1}\eea
The heat capacity
for the black plane can be straightforwardly computed using the fundamental Relation (\ref{bb1}) below:
\bea
{{C}_{Q}}=T{{(\frac{\partial S}{\partial T})}_{Q}}=\frac{2S(3{{\alpha }^{2}}{{S}^{2}}-\eta {{\pi }^{2}}{{Q}^{2}})}{3({{\alpha }^{2}}{{S}^{2}}+\eta {{\pi }^{2}}{{Q}^{2}})}\
\eea
Phase transitions are then determined by the roots of the denominator of $C_{Q}$; i.e., when the specific heat diverges. So, there exists only one divergent point $S_{i}=-i\sqrt{\eta }\pi Q/\alpha$,
which shows that the normal case ($\eta$=1) has no phase transition, while the phantom case ($\eta$=-1) possesses a phase transition.
For the phantom black plane, we compute the scalar curvature by using our formalism of the thermodynamic geometry. We can consider the following Legendre transformation of $M(S,Q)$:
\bea
\overline{M}(S,\Phi )=M(S,Q)-\Phi Q\
\label{kk2}\eea
Using (\ref{kk1}) and replacing Q by $\Phi \alpha \sqrt{\frac{S}{2{{\pi }^{2}}}}$ in (\ref{kk2}), we have:
\bea
\overline{M}(S,\Phi )=\frac{\alpha \sqrt{2S}(2S\eta -{{\Phi }^{2}})}{4\eta \pi }\
\eea
Now, we can evaluate the Ricci scalar $R(S,\Phi)$ for the black plane,
\bea
 \overline{R}(S,\Phi)=4/3\,{\frac {{\Phi}^{2} \left( 18\,S\eta+5\,{\Phi}^
{2} \right) \eta}{ \left( 2\,S\eta+{\Phi}^{2} \right) ^{2} \left( 6\,S
\eta-{\Phi}^{2} \right) }}
\eea
 By replacing Relation (\ref{kk1}) for $\Phi$, we have:
\bea
\overline{R}(S,Q)=\frac{-2{{\pi }^{2}}{{Q}^{2}}{{\alpha }^{2}}S\eta (9{{S}^{2}}{{\alpha }^{2}}+5\eta {{\pi }^{2}}{{Q}^{2}})}{{{({{S}^{2}}{{\alpha }^{2}}+\eta {{\pi }^{2}}{{Q}^{2}})}^{2}}(-3{{S}^{2}}{{\alpha }^{2}}+\eta {{\pi }^{2}}{{Q}^{2}})}\
\eea
The roots of the first term in  the denominator correspond to the phase transition points. The second factor in the denominator is zero only at the extremal limit ($T=0$).

\section{Thermodynamic geometry of Kerr Newman black hole}
Kerr Newman black holes \cite{R2} are described by their mass ($M$), entropy ($S$), charge ($Q$), and angular momentum ($J$). The mass formula for the Kerr Newman black hole is given by:
\bea
M=\frac{\sqrt{S(4{{J}^{2}}+{{S}^{2}}+2{{Q}^{2}}S+{{Q}^{4}})}}{2S}\
\label{w10}
\eea
The thermodynamic variables and the first law of thermodynamics are given by:
\bea
&&T=(\frac{\partial M}{\partial S})=\frac{{{S}^{2}}-4{{J}^{2}}-{{Q}^{2}}}{4{{S}^{\frac{3}{2}}}\sqrt{{{S}^{2}}+4{{J}^{2}}+{{Q}^{2}}+2{{Q}^{2}}S}}\\
&&\nonumber=\frac{{{S}^{2}}-4{{J}^{2}}-{{Q}^{2}}}{4{{S}^{2}}M}\
\eea
\bea
&&\Omega =(\frac{\partial M}{\partial J})=\frac{2J}{\sqrt{{{S}^{2}}+4{{J}^{2}}+{{Q}^{2}}+2{{Q}^{2}}S}\sqrt{S}}\\
&&\nonumber=\frac{J}{SM}\
\label{w13}
\eea
\bea
&&\Phi =(\frac{\partial M}{\partial Q})=\frac{4QS+4{{Q}^{3}}}{4\sqrt{S(4{{J}^{2}}+{{S}^{2}}+2{{Q}^{2}}S+{{Q}^{4}})}}\\
&&\nonumber=\frac{Q(S+{{Q}^{2}})}{2MS}\
\eea
\bea
dM=TdS+\Omega dJ+\Phi dQ\
\eea
where, $T$ is the Hawking temperature, $\Omega$ is the angular velocity, and  $\Phi$ is the potential deference of the Kerr Newman black holes.
The following relations also hold among the seven  variables $ M, T, S, \Omega, \Phi, Q, J$:
\bea
2T{{(\frac{1}{S}-{{\Omega }^{2}})}^{1/2}}=\frac{1}{2S}-{{\Omega }^{2}}-\frac{{{Q}^{2}}}{{{S}^{2}}}\
\eea
\bea
{{\Omega }^{2}}=\frac{1}{S}-{{(\frac{\Phi }{Q})}^{2}}\
\label{w11}
\eea
\bea
S(\frac{\Phi }{\Omega })=M+2TS\
\eea
These relations help us to achieve exact expressions for the heat capacities.

\noindent $C_{J,Q}$  is given by the following equation:
 \bea\label{w23}
 && \no {{C}_{J,Q}}=\frac{T}{{{(\frac{\partial T}{\partial S})}_{J,Q}}}=F(S,Q,J)\times\\  && \no1/(48{{J}^{4}}+24{{S}^{2}}{{J}^{2}}+32S{{J}^{2}}{{Q}^{2}}
+24{{J}^{2}}{{Q}^{4}}\\
 &&-{{S}^{4}}+6{{S}^{2}}{{Q}^{4}}+8S{{Q}^{6}})
 \eea
 The conjugate potential for $M(S,Q,J)$ can be defined as:
 \bea
 \overline{M}(S,\Omega ,\Phi )=M(S,Q,J)-\Omega J-\Phi Q\
 \label{w15}
 \eea
  Now, replacing $ J$ by  $(\Omega S M)$  in Equation (\ref{w10}) and resolving this with respect to $M$  will yield following relation for $M$:
 \bea
 M \left( S, \Omega, Q \right) =\,{\frac {S+{Q}^{2}}{2\sqrt {-{\Omega}^{2}{S}
^{2}+S}}}\
\label{w12}
\eea
 Using (\ref{w11}) and replacing $Q$ by ${\frac {\Phi\,\sqrt {S}}{\sqrt {1-{\Omega}^{2}S}}}$ in (\ref{w12}), we have,
 \bea
 M(S,\Omega, \Phi)=\frac{(S-S^{2}\Omega^{2}+S \Phi^{2})}{2(1-S \Omega^{2})\sqrt{S-S^{2}\Omega^{2}}}
 \eea
  Finally, we obtain the following conjugate potential:
 \bea
 \overline{M}(S,\Omega,\Phi)=\frac{S({{\Omega }^{2}}S-1)({{\Omega }^{2}}S-1+{{\Phi }^{2}})}{2(1-S{{\Omega }^{2}})\sqrt{S-{{S}^{2}}{{\Omega }^{2}}}}\
 \eea
  Defining the following metric:
   \beq
   {{\overline{g}}_{ij}}=\frac{1}{T}\left( \frac{{{\partial }^{2}}\overline{M}}{\partial {{X}^{i}}\partial {{X}^{j}}} \right);{{X}^{i}}=(S,\Omega ,\Phi )
   \eeq
 We may obtain  the resulting scalar curvature in terms of $S$, $\Omega$, and $\Phi$ as follows:
\bea
\overline{R}(S,\Omega,\Phi)=\frac{E(S,\Omega ,\Phi )}{A(S, \Omega, \Phi)^{2}B(S, \Omega, \Phi)}
\eea
We can also rewrite $C_{J,Q}$ as a function of $S$, $\Omega$, and $\Phi$ by replacing the following equations in Relation (\ref{w23}).

\bea
  J=\frac{\Omega S}{2}\left( \frac{S}{\sqrt{S-{{S}^{2}}{{\Omega }^{2}}}}+\frac{{{\Phi }^{2}}}{\sqrt{S-{{S}^{2}}{{\Omega }^{2}}}\left( {{S}^{-1}}-{{\Omega }^{2}} \right)} \right)
\eea
\bea
 Q=\frac {\Phi\,\sqrt {S}}{\sqrt {1-{\Omega}^{2}S}}\
\eea
Thus,
\begin{equation}
{{C}_{J,Q}}=\frac{2B(S,\Phi ,\Omega )\left( -1+{{\Omega }^{2}}S-{{\phi }^{2}} \right)}{A(S,\Phi ,\Omega )}
\end{equation}
where,
\bea
&&A=4{{\Omega }^{8}}{{S}^{4}}-16{{\Omega }^{6}}{{S}^{3}}+21{{\Omega }^{4}}{{S}^{2}}+\\
&&\nonumber4{{\Omega }^{4}}{{\Phi }^{2}}{{S}^{2}}-2{{\Omega }^{2}}{{\Phi }^{2}}S-10{{\Omega }^{2}}S+1-3{{\Phi }^{4}}-2{{\Phi }^{2}}
 \eea
   \bea
   B=2S(\Omega^{4}S^{2}-3\Omega^{2}S+1-\Phi^{2})
 \eea
 As a result, the roots of $A(S, \Omega, \Phi)$ correspond to the  divergence point of heat capacity $C_{J, Q}$, while $B(S, \Omega, \Phi)$ is zero only at the extreme points ($T=0$). For Myers-Perry black holes and a similar calculation see \cite{mirza3}.

\section{Conclusion}
In this work, we have explored a  formulation for the thermodynamic geometry of black holes. This formulation yields a proper expression of the  relation between heat capacity and curvature singularities. We also investigated a large class of black holes  in all of which the singularity of specific heat corresponds to that of scalar curvature.
We conclude that our method can be used as a correct and simple formulation for the characterization of the thermodynamic geometry of black holes and other thermodynamic systems.\\
\\
\\
\textbf{\large Appendix }

The first law of thermodynamics for a  black hole with the three parameters $S$, $Q$, and $J$ can be  written as follows:
\begin{equation}\label{q4}
dM=TdS+\Phi dQ+\Omega dJ
\end{equation}
The conjugate potential $\overline{M}(S,\Phi ,\Omega )$ can be obtained from $M(S,Q,J)$ by the following Legendre transformation:
\begin{equation}\label{q5}
\overline{M}(S,\Phi ,\Omega )=M(S,Q,J)-\Phi Q-\Omega J
\end{equation}
So, the first law of thermodynamic for this function would be:
\begin{equation}\label{q6}
d\overline{M}=TdS-Qd\Phi -Jd\Omega
\end{equation}
We can write the Maxwell relations from Eqs (\ref{q4},\ref{q6}) which are given by:

\bea\label{w1}
&&{{\left( \frac{\partial \Phi }{\partial J} \right)}_{S,Q}}={{\left( \frac{\partial \Omega }{\partial Q} \right)}_{S,J}};
{{\left( \frac{\partial \Phi }{\partial S} \right)}_{J,Q}}={{\left( \frac{\partial T}{\partial Q} \right)}_{J,S}}\\
&&\nonumber{{\left( \frac{\partial \Omega }{\partial S} \right)}_{J,Q}}={{\left( \frac{\partial T}{\partial J} \right)}_{S,Q}}
\eea
\bea\label{w2}
&&{{\left( \frac{\partial Q}{\partial \Omega } \right)}_{\Phi ,S}}={{\left( \frac{\partial J}{\partial \Phi } \right)}_{S,\Omega }};{{\left( \frac{\partial Q}{\partial S} \right)}_{\Omega ,\Phi }}=-{{\left( \frac{\partial T}{\partial \Phi } \right)}_{S,\Omega }}\\
&&\nonumber{{\left( \frac{\partial J}{\partial S} \right)}_{\Omega ,\Phi }}=-{{\left( \frac{\partial T}{\partial \Omega } \right)}_{S,\Phi }}
\eea
In addition, the Ruppeiner metric of the three parameters of black hole can be expressed by the following relations:
\begin{equation}\label{f1}
{{g}_{SS}}=\frac{1}{T}\left( \frac{{{\partial }^{2}}M}{\partial {{S}^{2}}} \right)=\frac{1}{T}{{\left( \frac{\partial T}{\partial S} \right)}_{J,Q}}
\end{equation}
\begin{equation}\label{f2}
{{g}_{SQ}}={{g}_{QS}}=\frac{1}{T}\left( \frac{{{\partial }^{2}}M}{\partial S\partial Q} \right)=\frac{1}{T}{{\left( \frac{\partial T}{\partial Q} \right)}_{S,J}}\\
\end{equation}
\begin{equation}\label{f3}
{{g}_{SJ}}={{g}_{JS}}=\frac{1}{T}\left( \frac{{{\partial }^{2}}M}{\partial S\partial J} \right)=\frac{1}{T}{{\left( \frac{\partial T}{\partial J} \right)}_{S,Q}}
\end{equation}
\begin{equation}\label{f4}
{{g}_{QJ}}={{g}_{JQ}}=\frac{1}{T}\left( \frac{{{\partial }^{2}}M}{\partial Q\partial J} \right)=\frac{1}{T}{{\left( \frac{\partial \Phi }{\partial J} \right)}_{S,Q}}
\end{equation}
\bea\label{f5}
{{g}_{QQ}}=\frac{1}{T}\left( \frac{{{\partial }^{2}}M}{\partial {{Q}^{2}}} \right)=\frac{1}{T}{{\left( \frac{\partial \Phi }{\partial Q} \right)}_{J,S}}
\eea
\bea\label{k10}
{{g}_{JJ}}=\frac{1}{T}\left( \frac{{{\partial }^{2}}M}{\partial {{J}^{2}}} \right)=\frac{1}{T}{{\left( \frac{\partial \Omega }{\partial J} \right)}_{S,Q}}
\eea
We can expand the right hand side of (\ref{q3}) as follows:
\bea\label{e3}
&&{{g}_{SS}}({{g}_{QQ}}{{g}_{JJ}}-{{({{g}_{QJ}})}^{2}})-{{g}_{SQ}}({{g}_{SQ}}{{g}_{JJ}}\\
&&\nonumber-{{g}_{SJ}}{{g}_{QJ}})+{{g}_{SJ}}({{g}_{SQ}}{{g}_{QJ}}-{{g}_{QQ}}{{g}_{SJ}})
\eea
By replacing the elements of metric (\ref{f1}-\ref{k10}) in (\ref{e3}) and using Maxwell relations (\ref{w1},\ref{w2}), we obtain  the following relation:
\bea
&&\left| \begin{array}{ccc}
\noalign{\medskip}   {{g}_{SS}} & {{g}_{SQ}} & {{g}_{SJ}}  \\
 \noalign{\medskip}  {{g}_{SQ}} & {{g}_{QQ}} & {{g}_{QJ}}  \\
 \noalign{\medskip}  {{g}_{SJ}} & {{g}_{QJ}} & {{g}_{JJ}}  \\
\end{array} \right|=\frac{{{({{C}_{\Phi ,\Omega }})}^{-1}}}{{{T}^{2}}}({{\left( \frac{\partial \Omega }{\partial J} \right)}_{S,Q}}{{\left( \frac{\partial \phi }{\partial Q} \right)}_{S,J}}\\
&&\nonumber-{{\left( \frac{\partial \Omega }{\partial Q} \right)}_{S,J}}{{\left( \frac{\partial \phi }{\partial J} \right)}_{S,Q}})
=\frac{{{({{C}_{\Omega \Phi }})}^{-1}}}{{{T}^{2}}}\left( \frac{\partial (\Omega ,\Phi )}{\partial {{(J,Q)}_{S}}} \right)
\eea

\end{document}